\magnification=\magstep1
\hfuzz=6pt
\baselineskip=15pt
\hsize= 6.0 truein

$ $
\vskip 1.5in
\centerline{{\bf A Greenberger-Horne-Zeilinger experiment 
for mixed states}
\footnote{$^\dagger$}{This work supported 
in part by
grant \# N00014-95-1-0975 from the Office of Naval
Research, and by ARO and DARPA under grant \#
DAAH04-96-1-0386 to
QUIC, the Quantum Information and Computation
initiative, and by a DARPA grant to 
NMRQC, the Nuclear Magnetic Resonance 
Quantum Computing initiative.}}

\bigskip
\centerline{Seth Lloyd}

\centerline{d'Arbeloff Laboratory for Information Sciences
and Technology}

\centerline{Department of Mechanical Engineering}

\centerline{MIT 3-160}

\centerline{Cambridge, Mass. 02139}

\bigskip\noindent{\it Abstract:} This paper shows how the
Greenberger-Horne-Zeilinger experiment, which demonstrates
the nonlocal nature of quantum mechanics, can be performed
using nuclear magnetic resonance on spins in
molecules at finite temperature.  
The use of nuclear magnetic resonance techniques
allows the experiment to uncover
the nonlocality not just of special GHZ states, but
of arbitrary three particle states.

\vfill\eject

Quantum mechanical systems exhibit nonlocal behavior that 
apparently violates our intuition about the classical 
world.$^{1-4}$
An example of such a nonlocal behavior is the 
Greenberger-Horne-Zeilinger effect,$^{2-3}$ which predicts that 
measurements made on three particles prepared in an 
entangled quantum state should give results that 
cannot be explained if each of the particles on its own 
is in some unknown classical state before the measurement.
This paper proposes a method for performing the GHZ experiment
using nuclear magnetic resonance (NMR).  

The following new results are derived.  First, even though
it is not possible to make macroscopic measurements on
individual spins, it is still possible
to perform the GHZ experiment using NMR by effectively
miniaturizing the data gathering and analysis device:
data is collected and analyzed using another spin on
the same molecule as the three spins in the GHZ state.
Second, in its NMR incarnation, 
the GHZ experiment can be performed at finite
temperature without initializing the spins in a pure
state: this paper provides
a procedure for performing experiments on {\it arbitrary
} three-spin states
that distinguish between the predictions of quantum 
mechanics and those of classical hidden variable theory.
Even at infinite temperature, in which the input state
is entirely mixed, it is still possible to use NMR to
perform an analog of the GHZ experiment.
These advances are of course qualified by the fact that
the three spins on which the experiment are performed
all sit in the same molecule together with the additional
spin that is used to gather and collate the data.  A
more satisfying version of the GHZ experiment would involve
performing the same experiment using widely separated
spins or photons.

In Mermin's version,$^3$ the GHZ experiment works as 
follows.  Consider three spin
$1/2$ particles in the state 
$$|\psi\rangle=(1/{\sqrt 2})\big(
|\uparrow^1_z\uparrow^2_z\uparrow^3_z\rangle
-|\downarrow^1_z\downarrow^2_z\downarrow^3_z\rangle\big)
\quad.\eqno(1)$$
\noindent Since $|\psi\rangle$ is an eigenstate of the operator 
the product operator 
$\sigma_x^1 \sigma_y^2 \sigma_y^3$ with eigenvalue $+1$,
a measurement of
the product of spin $x$ on the first spin, spin $y$ on 
the second spin, and spin $y$ on the third spin should
give the result $+1$.
Similarly,  
$|\psi\rangle$ is also an eigenstate of the operators 
$\sigma_y^1 \sigma_x^2 \sigma_y^3$ and 
$\sigma_y^1 \sigma_y^2 \sigma_x^3$ with eigenvalue $+1$,  
so that subsequent measurements
of the corresponding products should also yield the
result $+1$ while leaving the system in the state $|\psi\rangle$. 
Finally, it is also easily verified that the product of measurements 
of spin $x$ on all three spins gives the result $-1$:
$|\psi\rangle$ is an eigenstate of the operator $\sigma_x^1
\sigma_x^2 \sigma_x^3$ with eigenvalue $-1$.    

In summary, we have 
$$\eqalign{\sigma_x^1 &\sigma_y^2 \sigma_y^3|\psi\rangle= 
+1|\psi\rangle\cr 
\sigma_y^1 &\sigma_x^2 \sigma_y^3|\psi\rangle= +1|\psi\rangle\cr 
\sigma_y^1 &\sigma_y^2 \sigma_x^3|\psi\rangle= +1|\psi\rangle\cr 
\sigma_x^1 &\sigma_x^2 \sigma_x^3|\psi\rangle= -1|\psi\rangle
\quad.\cr }\eqno(2)$$ 
\noindent The contradiction with classical intuitions of locality
comes about as follows (a vivid description of this 
contradiction can be found in reference (3)).  
Suppose that there were an underlying
classical value (a so-called `hidden variable'),$^5$ 
$+1$ or $-1$, for each of the spins along 
both the $x$ and $y$ axes.  In this case, 
the measurements described in the previous paragraph 
could not yield the predicted result.  In particular, quantum mechanics
predicts that the product of the four product measurements 
spins should have the result $-1$.  But classically, each
measurement of a spin along a particular axis occurs twice
in the set of 12 measurements, and so the product of the
measurements should have the result $+1$.  The GHZ experiment
provides a natural framework for discriminating between the
predictions of quantum mechanics and the predictions of 
the simplest sort of classical hidden variable theory.
Unlike experiments involving Bell's inequality,$^6$ which give
a probabilistic method of descriminating between quantum 
and classical, the GHZ experiment need only be repeated once
in principle. 

A number of experiments have been proposed to verify the
GHZ effect but the difficulty of preparing the state
$|\psi\rangle$ and/or the difficulty of performing the individual
measurements has prevented these experiments from being realized.  
This paper proposes a method for performing the GHZ
experiment using using nuclear magnetic resonance.  

Let us for the moment ignore the difficulties of 
polarizing spins and
preparing the GHZ state.  Suppose that we have managed
to prepare four spins on each of a large number of
molecules in the state,
$$1/{\sqrt 2}\big( 
|\uparrow^1_z\rangle|\uparrow^2_z\rangle|\uparrow^3_z\rangle
-|\downarrow^1_z\rangle|\downarrow^2_z\rangle|\downarrow^3_z
\rangle \big)|\downarrow^4_z\rangle \quad.\eqno(3)$$
so that the first three spins are in the GHZ state 
and the fourth spin 
is initially in the state $|\downarrow^4_z\rangle$. 

Double resonance methods can now be used to make the
fourth spin perform the GHZ measurement.
Assume that the spins have different
resonant frequencies $\omega_i$, and
are coupled together according to the normal 
scalar interaction.  
Double resonance methods can now be used to
measure the various product operators.$^{7-8}$ 
$\sigma^1_x\sigma^2_y\sigma^3_z$, for example, one performs
a series of Controlled-NOT operations with the three spins
in the GHZ state as the controls.  (A thorough guide to
how to perform such quantum logic operations on spins is
given by Gershenfeld and Chuang.$^{9}$)
The sequence is as
follows:   

\bigskip
(0) Flip spin 4 into the state $|\uparrow^4_z\rangle$.

\bigskip
(1) Flip spin 4 if and only if spin 1 is in the state
$|\downarrow^1_x\rangle$.  (This operation can be
accomplished, for example, by rotating spin 1 by $\pi/2$ about
the $y$ axis so that $|\downarrow^1_x\rangle 
\longrightarrow |\uparrow^1_z\rangle$, performing a 
Controlled-NOT that flips spin 4 iff spin 1 is in the state
$|\uparrow^1_z\rangle$ and rotating spin 1 by $-\pi/2$ about
the $y$ axis to restore it to its original state.)

\bigskip
(2) Flip spin 4 if and only if spin 2 is in the state
$|\downarrow^2_y\rangle$.

\bigskip
(3) Flip spin 4 if and only if spin 3 is in the state
$|\downarrow^3_y\rangle$.

\bigskip
\noindent After this series of steps, it is clear that
spin 4 is in the state $|\uparrow^4_z\rangle$
if and only if it has been flipped an even
number of times.  That is, spin 4 now registers the product
of the results of measurements of spin $x$ on the first spin, spin
$y$ on the second spin, and spin $y$ on the third spin.
Quantum mechanics predicts that this sequence of
measurements should leave spin 4 in the state
$|\uparrow^4_z\rangle$.  In addition,
it is simple to verify that after steps (0-4), the first
three spins remain in the GHZ state.  That is, the steps make a
non-demolition measurement$^{10}$ of the product operator 
$\sigma^1_x \sigma^2_y \sigma^3_y$, and the GHZ 
state is an eigenstate of this operator.  

The prediction of quantum mechanics that the resulting
product is $+1$ can now be verified by tipping spin 4 by
$\pi/2$ about the $y$-axis in the co-rotating frame, and
by looking at the induction signal from the ensemble
of spins.  The result $+1$ corresponds to the induction
signal from spin 4 lying along the $x$-axis in the
co-rotating frame.  

To perform the full GHZ experiment, first measure
$\sigma^1_x\sigma^2_y\sigma^3_y$ as above, restore
spin 4 to the state $|\uparrow^4_z\rangle$ by tipping
the spin back by $-\pi/2$, then repeat steps
(1-3), suitably altering the conditional spin flip
operations to measure 
$\sigma^1_y\sigma^2_x\sigma^3_y$.  Once again, the
GHZ state is an eigenstate of this operator with
eigenvalue $+1$, so spin 4 should be flipped an even
number of times, leaving it in the state   
$|\uparrow^4_z\rangle$.  Repeat for 
$\sigma^1_y\sigma^2_y\sigma^3_x$ and for
$\sigma^1_x\sigma^2_x\sigma^3_x$, during which measurement
the spin is flipped an odd number of times and ends up
in the state $|\downarrow^4_z\rangle$.  The net result
of the sequential measurement of the four product operators,
according to quantum mechanics, is to flip spin 4, which may
be verified by rotating spin 4 by $\pi/2$ about the $y$-axis
in the co-rotating frame.  Since all of the many molecules
are performing the same measurement,  
quantum mechanics predicts that
the resulting induction signal should reveal spin 4 
to point along the
$-x$-axis in the co-rotating frame.   
A classical local hidden variable theory, in contrast,
predicts that spin 4 should point along the $+x$-axis
in the co-rotating frame.

The preceding method shows that one can perform the GHZ 
experiment using NMR despite the impossibility of measuring
the polarization of individual spins.  One simply has
Avogadro's number of molecules performing the GHZ
experiment individually, and then reporting back in parallel
whether or not the predictions of GHZ are verified.
Of course, this is a highly localized experiment:
the spins of the GHZ state
and the `apparatus' consisting of the spin that
collects and collates the data all sit on the same
molecule.  As a result, this experiment
can be used to confirm the quantum predictions of GHZ, but 
not to make delayed choice experiments to rule out
nonstandard interactions between the spins.

The only problem that remains is that of preparing the
system in the initial state (3). 
In fact, no sophisticated state preparation is necessary 
for performing the GHZ experiment.
As will now be shown, one can start with the four
spins in a thermal state at room temperature and perform
exactly the steps above, and still obtain an experiment
that discriminates between the predictions of quantum mechanics
and the predictions of a classical hidden variable theory.

At room temperature, the spins are in an almost completely
mixed state.  Let us approximate the state of the first
three spins as being completely mixed with density
matrix proportional to the identity matrix. 
This fully 
mixed state is an equal superposition of the GHZ state
$$1/{\sqrt 2}\big( 
|\uparrow^1_z\rangle|\uparrow^2_z\rangle|\uparrow^3_z\rangle
-|\downarrow^1_z\rangle|\downarrow^2_z\rangle|\downarrow^3_z
\rangle \big) \quad(+++-)\eqno(4.0)$$
\noindent together with the `alternative' GHZ states
$$1/{\sqrt 2}\big( 
|\uparrow^1_z\rangle|\uparrow^2_z\rangle|\downarrow^3_z\rangle
-|\downarrow^1_z\rangle|\downarrow^2_z\rangle|\uparrow^3_z
\rangle \big) \quad(--+-)\eqno(4.1)$$
$$1/{\sqrt 2}\big( 
|\uparrow^1_z\rangle|\downarrow^2_z\rangle|\downarrow^3_z\rangle
-|\downarrow^1_z\rangle|\uparrow^2_z\rangle|\uparrow^3_z
\rangle \big) \quad(+---) \eqno(4.2)$$
$$1/{\sqrt 2}\big( 
|\uparrow^1_z\rangle|\uparrow^2_z\rangle|\uparrow^3_z\rangle
+|\downarrow^1_z\rangle|\downarrow^2_z\rangle|\downarrow^3_z
\rangle \big) \quad(---+)\eqno(4.3)$$
$$1/{\sqrt 2}\big( 
|\uparrow^1_z\rangle|\downarrow^2_z\rangle|\downarrow^3_z\rangle
+|\downarrow^1_z\rangle|\uparrow^2_z\rangle|\uparrow^3_z
\rangle \big) \quad(-+++)\eqno(4.4)$$
$$1/{\sqrt 2}\big( 
|\uparrow^1_z\rangle|\uparrow^2_z\rangle|\downarrow^3_z\rangle
+|\downarrow^1_z\rangle|\downarrow^2_z\rangle|\uparrow^3_z
\rangle \big) \quad(++-+)\eqno(4.5)$$
$$1/{\sqrt 2}\big( 
|\uparrow^1_z\rangle|\downarrow^2_z\rangle|\uparrow^3_z\rangle
-|\downarrow^1_z\rangle|\uparrow^2_z\rangle|\downarrow^3_z
\rangle \big) \quad(-+--)\eqno(4.6)$$
$$1/{\sqrt 2}\big( 
|\uparrow^1_z\rangle|\downarrow^2_z\rangle|\uparrow^3_z\rangle
+|\downarrow^1_z\rangle|\uparrow^2_z\rangle|\downarrow^3_z
\rangle \big) \quad(+-++)\eqno(4.7)$$
\noindent Each of these eight states is an eigenstate of
the operators 
$\sigma^1_x\sigma^2_y\sigma^3_y$,
$\sigma^1_y\sigma^2_x\sigma^3_y$,
$\sigma^1_y\sigma^2_y\sigma^3_x$,
$\sigma^1_x\sigma^2_x\sigma^3_x$, 
and next to each state is indicated the
eigenvalue of that state with respect to these operators.
We see that quantum mechanics predicts that
each of these eight states, like the original
GHZ state, should give a product of the GHZ measurements
equal to $-1$, while classical mechanics predicts the
result $+1$.  Since these eight states form a
basis for the eight-dimensional Hilbert space of the three
spins, the NMR procedure should confirm the predictions
of quantum mechanics for {\it any} state, including a fully
mixed state. 

If the nuclear magnetic resonance procedure for
verifying GHZ is performed on a fully mixed state,
the intermediate measurements (of 
$\sigma^1_y\sigma^2_x\sigma^3_y$, for example)
will yield no net induction signal.
It is only for the measurement of the product
of all four operators in (2) that
quantum mechanics predicts 
the GHZ experiment will result in an induction signal
for the final spin that is oriented along the $-x$-axis,
while classical mechanics predicts that the signal
should be oriented along the $+x$-axis.  (At finite
temperature the induction signal for the fourth spin
is the result of the slight excess of spins in the ensemble
that start in the state $|\downarrow^4_z\rangle$ over
those that start in the state $|\uparrow^4_z\rangle$.) 

At first it might seem surprising that the GHZ experiment
does not need a special GHZ state to function.  The
explanation is that the NMR procedure for verifying GHZ
effectively performs a non-demolition measurement of
the product of all four operators in (2).  But 
$$(\sigma^1_x\sigma^2_x\sigma^3_x)
(\sigma^1_y\sigma^2_y\sigma^3_x)
(\sigma^1_y\sigma^2_x\sigma^3_y)
(\sigma^1_x\sigma^2_y\sigma^3_y) = -1. \eqno(5)$$ 
\noindent That is, the product of the four operators is
just minus the identity matrix: any input state has the
eigenvalue $-1$ with respect to this operator, so any
input state gives the output $-1$ for the measurement.

Quantum mechanics predicts that the sequence
of conditional spin flips required to perform the GHZ 
experiment using NMR gives the same result 
for any input state, and as a result
verifies the GHZ prediction not just on the 
state of the spins described by Greenberger, Horne,
and Zeilinger, but on a thermal state
of the spins.  Just as before, classical hidden
variable theory predicts the opposite result for 
exactly the same reasons as in references (3-4):
each hidden value for the state of a spin appears
twice in the product of equation (5), so that
the overall output should be $+1$.

This paper has demonstrated how the GHZ experiment can
be performed using nuclear magnetic resonance.  Although
it is not possible to make macroscopic measurements
of the polarization of individual spins, the GHZ
experiment can still be performed in parallel
on Avogadro's number of molecules at once by effectively
miniaturizing the data collection and analysis so
that the result of the experiment can be reported by the
induction signal from one spin on each molecule.
Of course, the method of performing the GHZ experiment
described here is less satisfactory than an experiment
in which the spins or photons that make up the GHZ
state can be widely separated at the time of the
measurement.  Indeed, the experimental method 
described can hardly be termed `non-local' as
all operations take place within a few angstroms
of eachother: there is no question of being
able to performed delayed choice experiments in such
a context.  The experiment described should be 
considered to be a confirmation of the non-local   
predictions of quantum mechanics within a local
experimental apparatus.  

There are two benefits of performing the measurement 
locally.  First, it allows the experiment to be
performed at all.  Second, the fact that measuring
apparatus and data analysis are performed by another
spin on the same molecule allows the non-demolition
measurement of the full product operator 
of equation (5).  As a result, the GHZ experiment 
can be performed on {\it any} input state, including 
thermal states.

\vfill
\noindent {\it Acknowledgements:} The author would like
to acknowledge helpful discussions with I. Chuang, 
N. Gershenfeld, and J.S. Waugh.
\eject

\centerline{\bf References}
\bigskip
\noindent 1. A. Einstein, B. Podolsy, N. Rosen, {\it Phys.
Rev.} {\bf 47}, 777-780 (1935).  This paper, together
with a number of seminal papers on quantum non-locality,
including references 2,5,6, and 7,
can be found in {\it Quantum Theory and Measurement},
J.A. Wheeler, W.H. Zurek, eds. (Princeton University Press,
Princeton, 1983).

\noindent 2. J.S. Bell, {\it Physics} {\bf 1}, 195-200
(1964).

\noindent 3. D.M. Greenberger, M. Horne, A. Zeilinger,
in {\it Bell's Theorem, Quantum Theory, and Conceptions
of the Universe}, M. Kafatos, ed. (Kluwer Academic,
Dordrecht, 1989), pp. 69-72. D.M. Greenberger, M.A.
Horne, A. Shimony, A. Zeilinger, {\it Am. J. Phys.}
{\bf 58}, 1131-1143 (1990).

\noindent 4. N.D. Mermin, {\it Am. J. Phys.} {\bf 58} (8),
731-734, (1990).

\noindent 5. D. Bohm, {\it Phys. Rev.} {\bf 85}, 166-193
(1952).  J.F. Clauser, M. Horne, {\it Phys. Rev. D} {\bf 10},
526-535 (1974).

\noindent 6. A. Aspect, {\it Phys. Rev. D} {\bf 14}, 1944-1951
(1976).

\noindent 7. C.P. Slichter, {\it Principles of Magnetic 
Resonance}, third edition (Springer-Verlag, New York, 1990).

\noindent 8. O.R. Ernst, G. Bodenhausen, Q. Wokaun, {\it
Principles of Nuclear Magnetic Resonance in One and Two
Dimensions} (Oxford University Press, Oxford, 1987).

\noindent 9. N.A. Gershenfeld, I.L. Chuang, {\it Science}
{\bf 275}, 350-356 (1997).
 
\noindent 10. C.M. Caves, K.S. Thorne, R.W.P. Drever, V.D.
Sandberg, M. Zimmerman, {\it Rev. Mod. Phys.} {\bf 52}, 341
(1980).

\vfill\eject\end